# Smart Sentiment Analysis-based Search Engine Classification Intelligence

1st Mike Nkongolo
*Faculty of Engineering, Built Environment & Information Technology*
*Department of Informatics, University of Pretoria*
Pretoria, South Africa
mike.wankongolo@up.ac.za

***Abstract*—Search engines are widely used for finding information on the Internet. However, there are limitations in the current search approach, such as providing popular but not necessarily relevant results. This research addresses the issue of polysemy in search results by implementing a search function that determines the sentimentality of the retrieved information. The study utilizes a web crawler to collect data from the British Broadcasting Corporation (BBC) news site, and the sentimentality of the news articles is determined using the Sentistrength program. The results demonstrate that the proposed search function improves recall value while accurately retrieving nonpolysemous news. Furthermore, Sentistrength outperforms deep learning and clustering methods in classifying search results. The methodology presented in this article can be applied to analyze the sentimentality and reputation of entities on the Internet.

## I. Introduction

Nowadays, popular news sites like China News and BBC News incorporate search functions, but these functions often lack advanced options like sentiment analysis. Users typically search for news using keywords, and search engines prioritize results based on relevance and popularity, rather than the user's specific expectations [1], [2]. This leads to a problem when a keyword has multiple meanings, as the desired results may not be obtained. While search engines like Google PageRank prioritize recent and reliable results, they may not always be relevant to the user's query [1]. For instance, a search for *apples* might yield results related to both fruits and phones, highlighting the issue of polysemy. To address this, users should be able to provide additional information to refine the search scope and improve accuracy. In this research, we focused on a problem that has received limited attention in existing articles. For instance, a previous study [3] manually collected search results and employed anomaly detection and neural networks for classification. In contrast, we implemented a search function on the search engine itself to improve the classification of search results. The proposed approach allowed for automatic and efficient data collection compared to the manual approach used in [3]. Furthermore, the method incorporates sentiment analysis, aiming to provide a framework that quickly retrieves desired internet data, classifies patterns, and assesses the online presence of entities. By applying the search function to BBC news and employing the Sentistrength algorithm, we were able to detect the polarity (negative, neutral, or positive) of the news content. This research investigates whether the classified search results impact the accuracy of the Sentistrength algorithm in sentiment analysis. We constructed a database containing diverse categories of news, although the search function was not capable of searching the entire Internet. Nonetheless, it collaborated with the search engine to gather data automatically. Each news article in the database was labeled, enabling the application of the Sentistrength computation. Furthermore, we implemented fixed tags that were categorized to enhance the organization of the data. The structure of this article consists of the background, which is discussed in Section II, followed by the research methodology presented in Section III. The results obtained are elaborated upon in Section IV, and finally, the conclusion and recommendations are provided in Section V.

## II. Background

The authors in [4] utilized deep learning and machine learning techniques to perform sentiment analysis on COVID19 data. The experimental results demonstrated that their proposed approach achieved an accuracy ranging from 93% to 95%, outperforming other techniques examined in the study. Another study conducted by [5] employed a Natural Language Processing algorithm to segment a collection of medical publications. In their methodology, both the words and the information contained within these segmented words were analyzed. The methodology focused on evaluating the effectiveness of the Natural Language Processing algorithm in the field of medicine. However, it should be noted that in this particular study, the algorithm was solely applied to process the content of headlines without utilizing pattern matching. The benefits of employing a search engine optimization approach were discussed in [6]. The study utilized Google to extract valuable information such as website browsing duration and user location, which are critical factors for optimizing search engine performance.

These features can be utilized to analyze user behavior and improve the overall search engine experience. In our methodology, when users input keywords on the search engine's homepage, only a predefined number of classified BBC news articles are displayed. By incorporating users' behaviors into our search function, it is possible for the search engine to make predictions about users' intended meanings when they enter polysemous keywords. For instance, if a user frequently searches for animal-related information, and they input the keyword *Jaguar*, the search engine can anticipate that the user is likely looking for information about the animal rather than the car. This represents the proposed optimal classification solution for addressing the polysemic problem discussed in this article. In a related study, [7] introduced an inverted index method that enables rapid execution of temporal queries. The applicability of this method was demonstrated using a COVID19 dataset. To enhance COVID-19 research, a novel approach called reverse sort-index was implemented, enabling real-time querying. This approach included various query categories, such as non-temporal, relative temporal, and absolute temporal. Experimental results demonstrated the effectiveness of the reverse sort indexing method compared to existing techniques, particularly in facilitating fast query execution for search engines. The inverted index proved to be efficient in retrieving nonpolysemic big data. In our methodology, we focused on searching and storing 800 BBC news articles in the database. However, the proposed method can be further applied to other search engines to extract a larger volume of data. It is worth noting that sorting indexes, as discussed in [7], is also an important aspect to consider. In their study, Xu et al. [8] developed a text-driven framework for aircraft fault diagnosis. The framework incorporated Word2vec and Convolution Neural Network (CNN) techniques. Numerous text files were utilized in the experiments, where Word2vec was employed to retrieve textual pattern vectors. These vectors were then passed to the CNN, which made the final decision regarding aircraft fault diagnosis. The CNN model incorporated a Cloud Similarity Measure (CSM) to enhance the performance of the classifier and provide support for aircraft maintenance. By combining unstructured and structured patterns, [8] successfully determined the cause of the aircraft fault. In our approach, we employed Word2vector to extract the content of search results. Based on the similarity of vectors, we recommended search results to users. By employing this approach, users have the ability to view search results based on the keywords they entered and the similarity of the search results. Many existing search engines utilize similar methodologies but with variations in their optimization techniques. For instance, Reddy et al. [5] applied Natural Language Processing to extract nonpolysemic information from a vast collection of medical data. Similarly, in our research, we utilized Natural Language Processing to select relevant information from BBC news. The significance of search engine optimization was emphasized in Pawade et al. [6], where they employed an inverted index search method on a large dataset to enhance search engine performance. Similarly, to the approach described in [6], we have employed an inverted index technique in our research. However, our implementation consists of a smaller number of entries compared to the study mentioned. This highlights the potential of our search function to be applied in searching for large volumes of data.

## III. Research Methodology

To enhance the accuracy of the search function and address the problem of polysemy, a search feature was developed specifically for categorizing and identifying BBC news based on user-defined categories. In order to collect the necessary data, a web mining methodology was employed to extract the content from web documents. The data retrieval process involved crawling the BBC news website using a web crawler, as described in [9]. Following the pre-processing and inverted index stages using the NLTK toolkit, the obtained data was stored in a database for further analysis.

### *A. The BBC data*

Due to its status as the largest news broadcaster in the world [10], we chose to crawl data from the BBC website, which ensured the reliability and credibility of the news content examined in this research [11], [10]. Furthermore, the BBC news website covers a wide range of topics, making it well-suited for our search function. For our data sample, we randomly selected 800 BBC news articles, ensuring a stochastic selection process to maintain the reliability of the search results. These articles were stored in a database table called *news*, containing fields such as the Uniform Resource Locator (URL), content, date (dt), title, and label. The label field indicates the classified information associated with each BBC news article. Finally, we performed sentiment analysis on the news table to determine the overall tone of the content, which can be valuable in assessing the influence and impact of BBC news on the web.

### *B. Crawling*

Web crawling involves the automated search for information on the Internet using predefined rules or conditions. In our research, we employ a web crawler due to its numerous advantages in efficiently downloading web pages and enabling multi-threading. For web crawling, we utilize a Python-based Google App Engine (PGAE), as introduced in [12]. The configuration container includes a core file that encompasses essential parameters such as the interval time between two fetches, the directory path for storing the retrieved data, and the designated start and end times for the crawling process.

The interval time between two fetches regulates the frequency at which the website is crawled to retrieve data. If the crawling frequency is set too high, there is a risk of the Internet Protocol (IP) address being blocked. In the data engineering process, we made use of HyperText Markup Language (HTML) div and span tags to identify the structure of the documents. Additionally, HTML components such as paragraph (p) and emphasis (em) were utilized to accurately represent the semantics of the web content. However, the inclusion of div and span tags can hinder the accessibility of the web content. To address this, we cleaned the crawled features by removing any div tags from the web pages. The resulting cleaned data was then stored in a JSON file, which will subsequently be transferred and stored in a database in the form of a table. In this table (news), the headers of the web pages were defined to store the BBC news. Refer to Fig. 1 for an illustration of the table storing the BBC news.

| id | label | url | title | dt | article |
|---|---|---|---|---|---|
| 11 | London | http://www.bbc.co.uk/news/ | Sarah Everard disappearance: Met officer arrested o | 2021-03-10 | A serving Met police officer has been arre |
| 12 | London | http://www.bbc.co.uk/news/ | Sarah Everard: Human remains found in Kent woodl | 2021-03-10 | Human remains have been found in the se |
| 13 | Scotland | http://www.bbc.co.uk/news/ | Covid in Scotland: Rules on people meeting outdoo | 2021-03-10 | Up to four adults from two different hous |
| 14 | UK Politics | http://www.bbc.co.uk/news/ | PMQs: As it happened - PM challenged on NHS pay | 2021-03-10 | Amid the blizzard of figures from the prin |
| 15 | UK Politics | http://www.bbc.co.uk/news/ | Act now on LGBT+ conversion therapy, ministers urg | 2021-03-10 | LGBT+ campaigners have urged the gover |
| 16 | Business | http://www.bbc.co.uk/news/ | Heathrow says airport border queues at 'unacceptal | 2021-03-10 | Heathrow Airport regularly sees queues o |
| 17 | UK | http://www.bbc.co.uk/news/ | Sellafield nuclear site a 'toxic mix of bullying and ha | 2021-03-10 | A "toxic culture" of bullying and harassme |
| 18 | Newsbeat | http://www.bbc.co.uk/news/ | Breonna Taylor's boyfriend Kenneth Walker cleared | 2021-03-10 | Kenneth Walker said he thought an intruc |
| 19 | Wales | http://www.bbc.co.uk/news/ | Covid: 'Stay-local' rules likely to differ across Wales | 2021-03-10 | Any "stay-local" policy introduced as lock |
| 20 | Australia | http://www.bbc.co.uk/news/ | Richard Pusey: Australian admits filming taunts of d | 2021-03-10 | An Australian man has pleaded guilty to |
| 21 | London | http://www.bbc.co.uk/news/ | Sarah Everard: Block of flats cordoned off in search f | 2021-03-10 | Officers searching for a missing woman in |
| 22 | Scotland | http://www.bbc.co.uk/news/ | Covid in Scotland: FM quizzed on lockdown easing | 2021-03-10 | 'We kid ourselves' if think vaccinating just |
| 23 | UK | http://www.bbc.co.uk/news/ | Meghan racism row: Society of Editors boss Ian Mur | 2021-03-10 | The executive director of an industry body |
| 24 | Asia | http://www.bbc.co.uk/news/ | Myanmar coup: 'We were told to shoot protesters', | 2021-03-10 | Myanmar coup: 'We were told to shoot pr |
| 25 | Liverpool | http://www.bbc.co.uk/news/ | Everton goalkeeper Robin Olsen and family in mach | 2021-03-10 | Everton goalkeeper Robin Olsen and his f |
| 26 | Wales | http://www.bbc.co.uk/news/ | Stalking protection orders: Police 'not using' new pc | 2021-03-10 | Stalking protection orders: Police 'not usi |
| 27 | UK Politics | http://www.bbc.co.uk/news/ | Air passenger duty: Review planned to cut tax on dc | 2021-03-10 | Boris Johnson has said he wants to cut air |
| 28 | UK Politics | http://www.bbc.co.uk/news/ | Boris Johnson accused of lying to MPs over Labour's | 2021-03-10 | Labour has accused Boris Johnson of lying |
| 29 | Wales | http://www.bbc.co.uk/news/ | Rhondda: Man, 31, charged with murder of girl, 16 | 2021-03-10 | Wenjing Lin died at the Blue Sky Chinese |
| 30 | Business | http://www.bbc.co.uk/news/ | Lego plans hiring spree for digital growth drive - BB | 2021-03-10 | Mr Christiansen's favourite Lego is the bu |
| 31 | Wiltshire | http://www.bbc.co.uk/news/ | Salisbury Novichok poisoning house to be bought E | 2021-03-10 | The house at the centre of the Salisbury N |
| 32 | Scotland | http://www.bbc.co.uk/news/ | Covid in Scotland: Sturgeon objective 'to eliminate' | 2021-03-10 | First Minister Nicola Sturgeon has said Sc |
| 33 | UK Politics | http://www.bbc.co.uk/news/ | Covid-19: UK hasn't banned export of vaccines, says | 2021-03-10 | Boris Johnson has "corrected" European C |
| 34 | London | http://www.bbc.co.uk/news/ | Sarah Everard disappearance: House and woods in K | 2021-03-10 | Police investigating the disappearance of |
| 35 | UK | http://www.bbc.co.uk/news/ | Meghan and Harry interview: Palace taking race issu | 2021-03-10 | The race issues raised by the Duke and Du |
| 36 | London | http://www.bbc.co.uk/news/ | Sarah Everard: Police confirm last sighting of missin | 2021-03-10 | The last confirmed sighting of Ms Everard |
| 37 | UK Politics | http://www.bbc.co.uk/news/ | SNP chief whip steps aside following harassment co | 2021-03-10 | Patrick Grady became his party's chief wh |
| 38 | UK | http://www.bbc.co.uk/news/ | Census 2021: Judge orders change to sex question | 2021-03-10 | As well as their legal sex, the census will a |
| 39 | London | http://www.bbc.co.uk/news/ | Sarah Everard: Family 'desperate' to see missing wo | 2021-03-10 | Detectives have asked people in the area |
| 40 | Business | http://www.bbc.co.uk/news/ | 'Right to repair' law to come in this summer - BBC N | 2021-03-10 | 'Right to repair' law to come in this summ |

Fig. 1. The BBC news table

*C. Search function implementation*

We chose the NLTK (Natural Language Toolkit) to implement the search function, which involved tokenizing sentences into individual words. By utilizing the inverted index, we were able to enhance the retrieval of pertinent information. Additionally, the Word2vec model was employed to compute the similarity between news articles. The Sentistrength algorithm was applied to perform sentiment analysis on the collected data, as described in [13].

*D. The NLTK and inverted index*

In this study, the Python NLTK library was utilized, specifically employing two methods: NLTK tokenize and NLTK stem. The NLTK tokenize function was employed to segment sentences into individual words, with spaces serving as the delimiters. Similarly, the NLTK stem function was applied to normalize words by transforming them into an acceptable format, such as converting past tense to present tense. For the purpose of this research, the NLTK tool was used to tokenize the search queries entered by users in the search engine. The proposed search function then performs searches based on the tokenized words derived from the search results. During the data cleaning phase, irrelevant words, known as stop words, were eliminated from the dataset. Stop words are commonly occurring words in a language that have little semantic significance. In this study, a list of six common determiners (a, that, the, an, and, those) was identified as irrelevant words in the web text. These determiners assist in describing nouns and expressing concepts related to localization or numbers. To enhance computational efficiency, stop words were removed from the dataset. By eliminating stop words, the number of indexes in the corpus was reduced, resulting in improved retrieval efficiency. Ignoring stop words in this research aimed to enhance the accuracy of the search function. For example, when searching for *apples*, the search engine may display 100 search results. However, if the search query includes *bananas* and *apples*, the search function would tokenize the sentence into three parts: (banana), (and), and (apples). Without removing the stop word, this approach could lead to errors and potentially retrieve more than 100 search results.

To improve the accuracy of computation, stop words were removed from the dataset in this research. The Python library used in the implementation of the search function included pretrained models and corpora. Inverted indexing was adopted to facilitate information retrieval based on attribute values. Each entry in the index table contained a specific attribute and the address of records with that attribute value. The position of the record was determined by its attribute value. An inverted index file was computed and created, based on tokenizing and cleaning the words from the BBC article using NLTK. These words were then indexed and stored in the database for sentiment analysis. When a user searches for these words, the index allows for rapid retrieval of relevant search results. The table depicted in Fig. 2 illustrates the storage of these indexes.

| term | df | docs |
|---|---|---|
| sarah | 7 | [11, 1, 10][12, 1, 9][21, 1, 10][34, 1, 9][36, 1, 10][39, 1, 9][42, 1, 8] |
| everard | 7 | [11, 1, 10][12, 1, 9][21, 1, 10][34, 1, 9][36, 1, 10][39, 1, 9][42, 1, 8] |
| disappearance | 2 | [11, 1, 10][34, 1, 9] |
| met | 7 | [11, 1, 10][329, 1, 11][338, 1, 11][443, 1, 12][502, 1, 10][628, 1, 8][702, 1, 13] |
| officer | 10 | [11, 1, 10][18, 1, 12][293, 1, 8][342, 1, 11][347, 1, 10][369, 1, 10][382, 1, 12][59 |
| arrested | 5 | [11, 1, 10][315, 1, 10][368, 1, 8][599, 1, 10][632, 1, 10] |
| suspicion | 1 | [11, 1, 10] |
| murder | 15 | [11, 1, 10][29, 1, 7][99, 1, 9][114, 1, 11][135, 1, 9][235, 1, 9][285, 1, 11][304, 1, |
| bbc | 697 | [11, 1, 10][12, 1, 9][13, 1, 9][14, 1, 8][15, 1, 8][16, 1, 9][17, 1, 9][18, 1, 12][19, 1, |
| news | 697 | [11, 1, 10][12, 1, 9][13, 1, 9][14, 1, 8][15, 1, 8][16, 1, 9][17, 1, 9][18, 1, 12][19, 1, |
| human | 1 | [12, 1, 9] |
| remains | 1 | [12, 1, 9] |
| found | 11 | [12, 1, 9][53, 1, 11][93, 1, 9][157, 1, 8][386, 1, 7][391, 1, 11][412, 1, 10][468, 1, |
| kent | 3 | [12, 1, 9][34, 1, 9][99, 1, 9] |
| woodland | 1 | [12, 1, 9] |
| covid | 100 | [13, 1, 9][19, 1, 9][22, 1, 8][32, 1, 9][47, 1, 7][58, 1, 8][61, 1, 10][72, 1, 10][78, 1, |
| scotland | 17 | [13, 1, 9][22, 1, 8][32, 1, 9][121, 1, 9][153, 1, 10][161, 1, 8][192, 1, 8][367, 1, 9][ |
| rules | 6 | [13, 1, 9][19, 1, 9][71, 1, 9][123, 1, 9][321, 1, 11][669, 1, 12] |
| people | 11 | [13, 1, 9][134, 1, 11][245, 1, 9][346, 1, 8][363, 1, 9][370, 1, 7][484, 1, 10][510, 1, |
| meeting | 1 | [13, 1, 9] |
| outdoors | 1 | [13, 1, 9] |
| eased | 2 | [13, 1, 9][321, 1, 11] |
| pmqs | 3 | [14, 1, 8][428, 1, 9][661, 1, 9] |
| happened | 23 | [14, 1, 8][61, 1, 10][98, 1, 6][155, 1, 8][168, 1, 10][202, 1, 7][246, 1, 8][270, 1, 1 |

Fig. 2. The database inverted index table

In the table shown in Fig. 2, the first column stores the search keywords entered by the user. The second column indicates the number of indexes that match the keywords, and the third column stores the specific keywords that the index corresponds to. For example, if a user searches for BBC, the table would display 697 indexes and articles that match that keyword. This implementation of inverted indexing enhances concurrency and automates the generation of attribute values, thereby determining the location of the corresponding records.

*E. The BM25 and Sentistrength algorithms*

The BM25 technique is a ranking method commonly used by search engines to estimate the relevance of a document based on a specific search query. In this approach, the user's keywords are tokenized into individual words, and these words are matched with the index file to determine their occurrences. A score is then calculated to measure the similarity of each web article to the search query. The search results are subsequently sorted based on this score.

Compared to the traditional Term Frequency Inverse Document Frequency (TFIDF) approach, BM25 incorporates adjustable parameters that enhance its power and flexibility. These parameters allow for more fine-tuning of the ranking process, resulting in improved search accuracy and effectiveness. The BM25 algorithm incorporates the concept of average document length, which considers the impact of a document based on its average length compared to other documents. Sentistrength, on the other hand, is a sentiment strength detection program that utilizes nonlexical linguistic rules and information.

Previous experiments have demonstrated the reliable performance of Sentistrength in sentiment analysis for web mining tasks. In the Sentistrength approach, each textual pattern is assigned three scores: negative, positive, and neutral. The overall polarity of the textual data is then computed by considering the emotional valence indicated by these scores. Equation 1 represents the calculation of the overall polarity.

$$polarity = \frac{positive + neutral + negative}{3} \quad (1)$$

Consequently, by averaging the polarity scores on a scale ranging from -5 (indicating negativity) to 5 (indicating positivity), the final polarity of the text can be determined. A score of 0 corresponds to a neutral polarity assignment. Fig. 3 illustrates the framework of our search function.

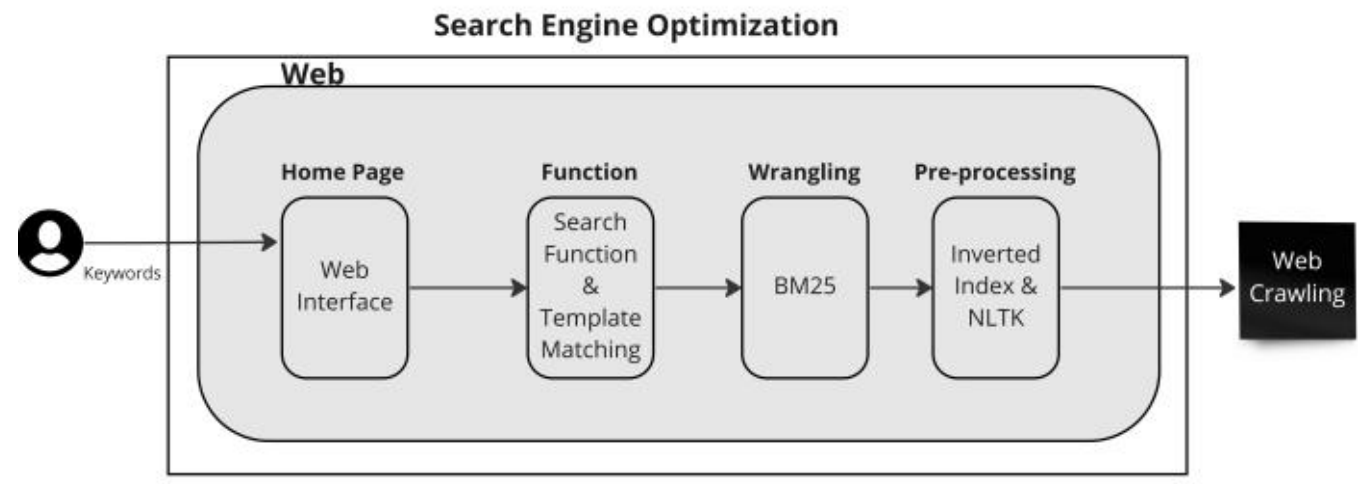

Fig. 3. The search function framework

The NLTK is utilized to tokenize the BBC news into words, where spaces are used as separators. Punctuation and full stops are removed using Regular Expressions, and all words are converted to lowercase. Irrelevant words with no meaning are eliminated, and word normalization is performed through stemming. The resulting words are restricted to their primitive form to ensure complete search functionality. The inverted index enables quick and efficient keyword searches, as each cleaned word has an index that corresponds to specific BBC articles.

The search results are ranked using the BM25 algorithm, which calculates scores based on the similarities between keywords and web articles. The search function module searches for users' keywords in the index table and retrieves the corresponding indexes to find matching news. In cases where entered keywords are not found in the index table, a Regular Expression matching process is applied to identify the most similar indexes and words. The interface module serves as the front end of the framework, displaying the search results to users based on their input keywords. The content page is used to present the search results to the user. Fig. 4 illustrates the process of the search function, which incorporates sentiment analysis.

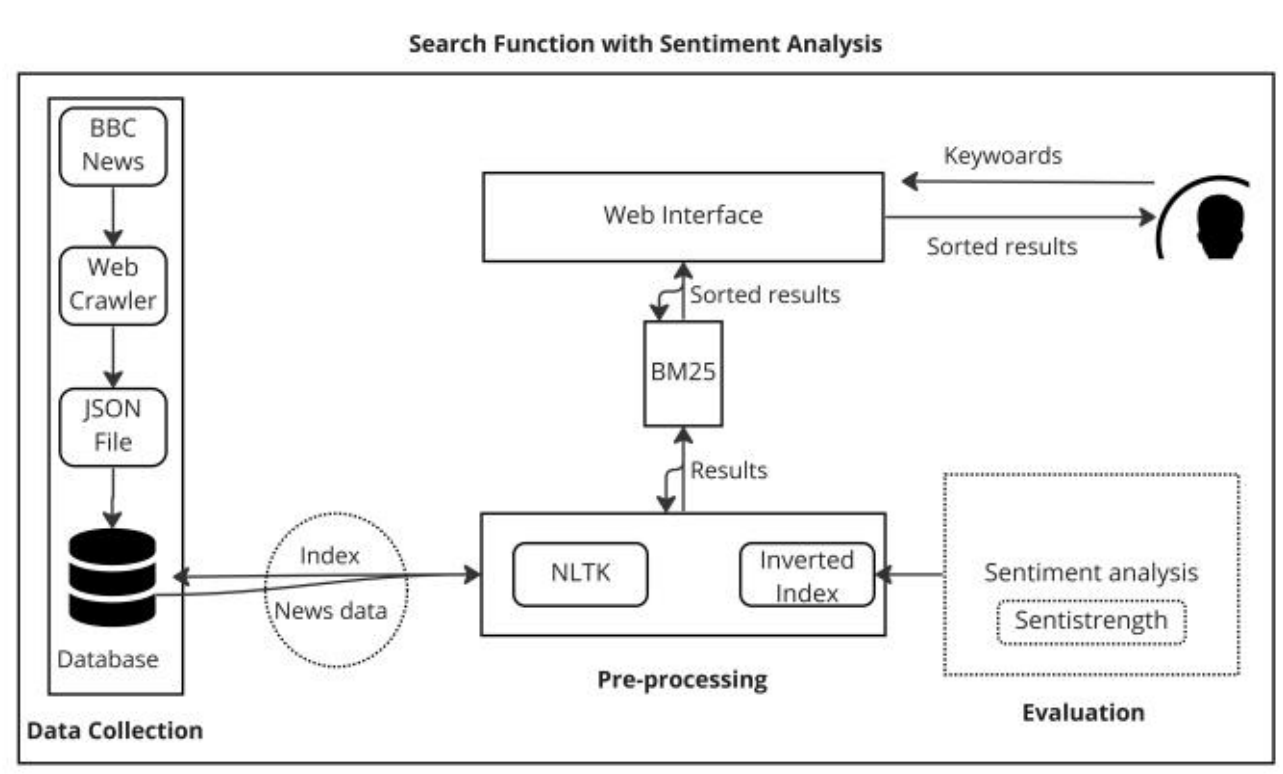

Fig. 4. Search function and sentiment analysis process

The data collection phase involves using a web crawler to gather news from the BBC website, which is then stored in a JSON file and transferred to the database.

The pre-processing module automates the extraction of relevant patterns from the stored news data in the database. This includes tokenizing the news features into words using the NLTK, removing punctuation and converting to lowercase, removing stop words, and normalizing the text through stemming. The cleaned words are then converted into indexes that represent the matched BBC articles, and these indexes are stored in a posting table in the database. During the search process, users input keywords through the web interface, which are searched against the database index table. The matching keywords are then used to retrieve the corresponding web articles, and in some cases, the results are also passed to the BM25 tool. The sorted search results are displayed on the web page for the users to see. Lastly, sentiment analysis is implemented on the database to evaluate the polarity of the results associated with the input keywords.

### *G. Evaluation metrics*

The search method is evaluated by comparing the proposed search function with a normal search function. Recall and precision rates are computed to assess the performance of the proposed search function. The sentiment analysis is also evaluated using the Sentistrength algorithm and BBC news data, with recall and precision metrics [14]. Finally, the recall and precision rates of both search functions are compared to the performance of Sentistrength. Precision (P) is calculated as the percentage of relevant news (RN) among the total number of retrieved news (TRN), as shown in Equation 2.

$$P = \frac{RN}{TRN} \times 100 \tag{2}$$

The recall (R) is a metric to measure the importance of the data retrieval module [15]. It is expressed in Equation 3.

$$R = \frac{RN}{TNS} \times 100 \tag{3}$$

Where TNS denotes the total amount of relevant news.

### *B. Experimental results*

In this section, we will calculate the recall and precision rates for the data using both the normal and classified search functions (Fig. 12). Additionally, we evaluate the performance of the Sentistrength algorithm using the same metrics [3], [16]. Furthermore, we compare the obtained results with existing methodologies, such as deep learning and clustering. The following section provides a detailed analysis of the recall and precision rates of the classified search function in comparison to the normal search (Fig. 12). The Sentistrength algorithm will be evaluated using the normal and classified search results, utilizing the same metrics.

Furthermore, a comparative analysis will be conducted, considering the results obtained from deep learning Fig. 8. Searching for Covid vaccine news and clustering methodologies. In Fig. 12, the classified search function achieved a recall rate of 99% and a precision of 50% for the Covid category. For the vaccine category, the classified search function achieved a recall rate of 90% and a precision of 33.8%. Additionally, in the travel category, the recall rate of the search function was 95% with a precision of 45%. These experimental findings highlight that the classified search function improves the recall rate by sacrificing precision. Therefore, the proposed search function approach has the potential to enhance the recall rate for internet news searches. It should be noted that polysemous words were utilized to test the search function, as they possess different meanings in various contexts, such as vaccine and travel. The classified search function effectively addresses the challenge of polysemy. The sentiment analysis computation, based on the data presented in Fig. 12, is visualized in Fig. 13. The results indicate a neutral polarity for BBC news (Fig. 13). Moreover, the Sentistrength algorithm significantly improves the recall and precision rates of the classified search function, achieving 100% and 75% respectively. In comparison, deep learning and clustering methods achieved a precision of 100%.

## V. Conclusion and discussion

This study focuses on improving the precision of the classified search function using the Sentistrength program. Although the experiments were conducted on a single database, the results can serve as a reference for search engine optimization studies. The proposed search function includes options for users to select different categories based on their specific news preferences. However, there are limitations to consider. The methodology cannot filter out fake news sites from the search results, and the number of categories may not cover all relevant groups for users. Expanding the list of categories can enhance search efficiency. Additionally, the search function algorithm can be enhanced by incorporating advanced classification schemes and intelligent tagging using machine learning. Future work could involve the automatic creation of search categories using artificial intelligence.

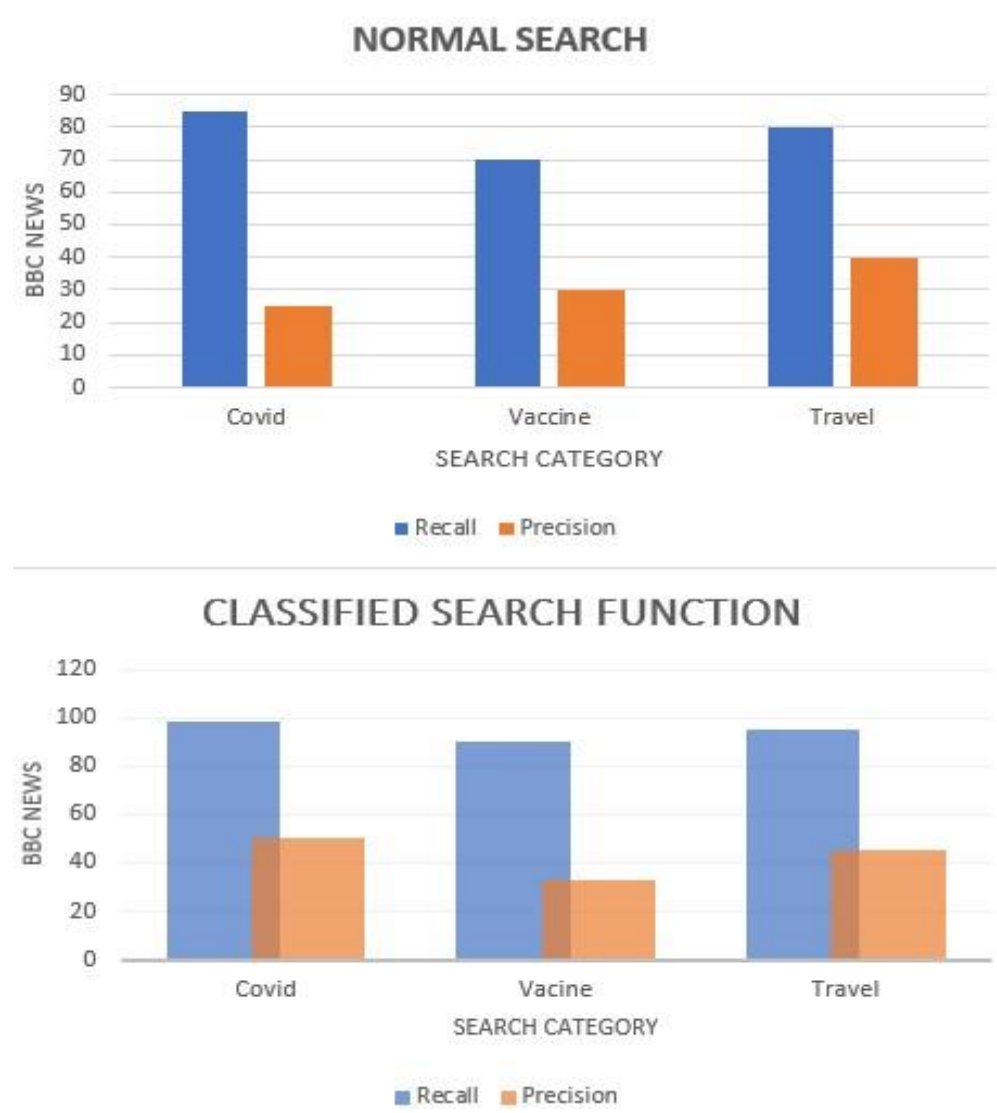

Fig. 12. The recall and precision rate of the search function

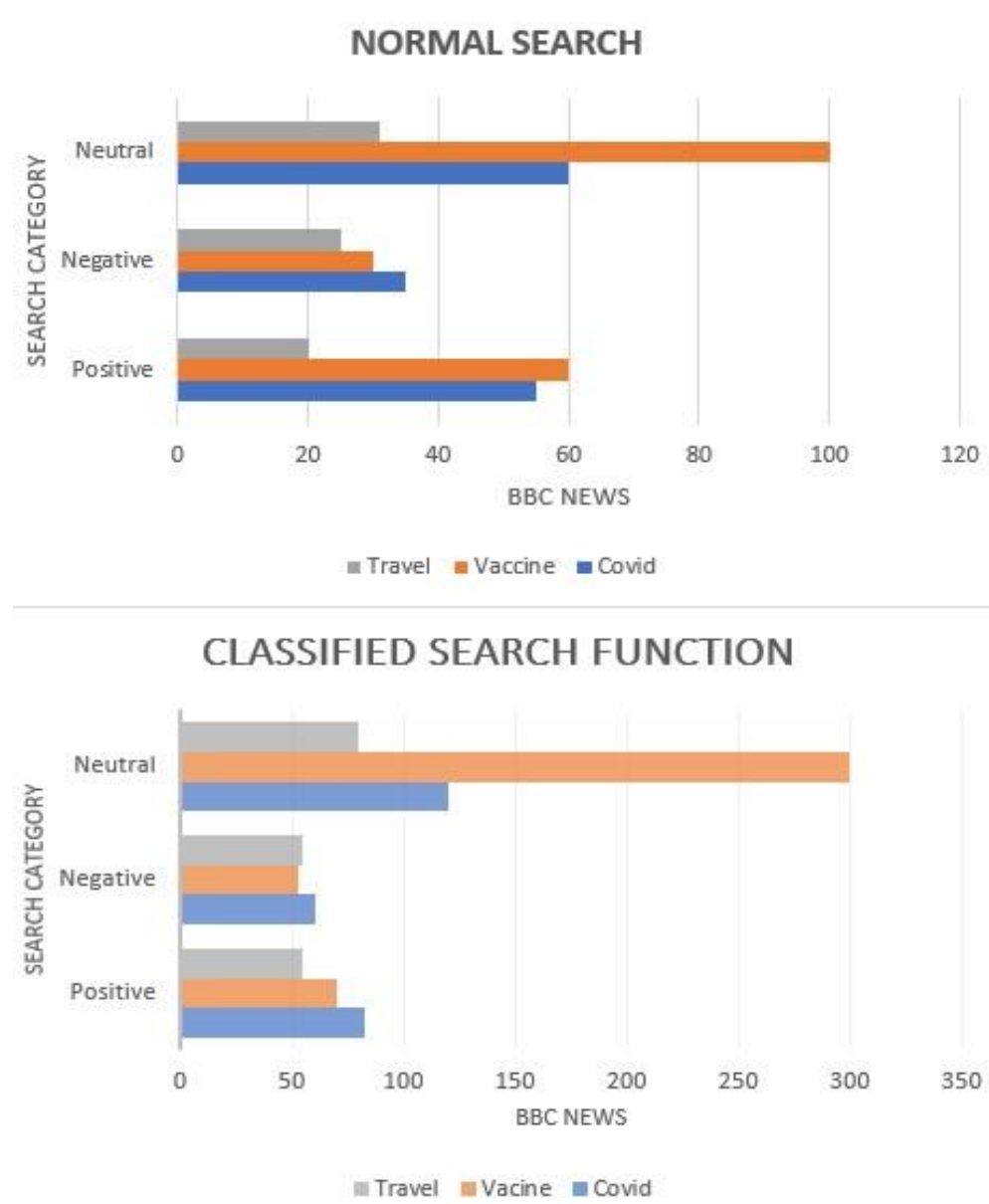

Fig. 13. The sentiment analysis computation